\documentclass[10pt,a4paper,twocolumn,byrevtex,nofootinbib,longbibliography,balancelastpage,showkeys]{revtex4-1}
\usepackage[utf8]{inputenc}
\usepackage{graphicx}
\usepackage{hyperref}
\usepackage{siunitx}

\AtBeginDocument{\usepackage{booktabs}}
\begin{document}
\title{Comment to ``Impact of Daylight Saving Time on circadian timing system: An expert statement''}
\author{José María Martín-Olalla}
\email{olalla@us.es}
\affiliation{Universidad de Sevilla\\Facultad de Física\\Departamento de Física de la Materia Condensada\\ES41012 Sevilla, Spain}
\date{\today}
\begin{abstract}
  Published in \emph{European Journal of Internal Medicine} (2019) doi: \href{https://doi.org/10.1016/j.ejim.2019.02.006}{10.1016/j.ejim.2019.02.006}
\end{abstract}

\maketitle

The impact of Daylight Saving Time (DST) on daily life has gained a lot of attention in the past few years and, notably, in the past months since the European Commission addressed the issue. It was suggested recently\cite{MeiraeCruz2019} through evidences that \emph{the impact of one-hour change of time} in the circadian timing system may be relevant in terms of public health. Perhaps inadvertently, authors only addressed one leg of the problem ---the magnitude of the change of time---, nonetheless they arrived to a bold statement ``DST should be discontinued.''

DST is not just a one-hour change of time. It is not randomly onset. It is not randomly offset. The direction of the change ---spring forward, fall back--- is not randomly chosen.  It is not randomly distributed on Earth.

DST is the way in which many contemporary societies have successfully addressed one specific issue: the seasonal variation of the phase of human activity. The root of the problem can be expressed as follows:  in winter the phase of the sleep/wake cycle delays with increasing latitude at the same rate sunrise does. The midpoint of wakefulness in pre-industrial, Subtropical societies\cite{Moreno2015,Yetish2015} delays one hour from noon. In Great Britain the lag increases in winter two more hours, up to three hours from noon. Winter sunrise also delays two hours from the Equator to Britain.\cite{Martin-Olalla2018b} Can this lag sustain in the opposite season?

I will try to outline an answer to this question confronting two opposing ideas: (1) sleep/wake cycle should not seasonally change; and (2) contemporary societies can achieve some degree of seasonality just following the evolution of sunrises and sunsets.

The second idea ceased to be an option after time keepers evolved from [seasonal] sundials to [nonseasonal] mechanical clocks, time was standardized, and the quantum of time change was set to one hour in practical terms. In fact, as early as in 1810, the Cádiz Cortes (first Spanish national assembly) already advanced by one hour the timing of their sitting from May to October. Later, on 1905, William Willet, one of the many fathers of DST, campaigned for advancing clocks in spring in four strokes of twenty minutes each. It was unpractical. Contemporary societies can not follow steadily seasons. Instead many of them wait until the Sun rises early enough ---spring--- to trigger the one-hour change in social timing.  In summary point (2) just ends in DST.

   The first idea somewhat couples the circadian timing system and human activity to clock time and Earth's rotation, which are nonseasonal. Therefore, no matter the fact the planet is tilted by \ang{23.5}, no matter the fact that in extratropical latitudes the Sun apparently rises in the sky $\ang{47}=2\times\ang{23.5}$ from winter to summer, no matter the changes in the light and dark (LD) cycle this triggers, social timing should not change year round.

   Below some circle of latitude, say for instance at Subtropical latitudes, this idea is quite obvious because natural changes in the LD cycle are quite small. DST was never a preference here.

    For opposite reasons, and somewhat paradoxically, this idea may also be sound above some circle of latitude when seasonal variations are so large that societies can not follow them. Winter sunrise comes too late, summer sunrise comes too early.

    Societies in this range could be struggling to reduce their lag by starting human activity well before sunrise in winter\cite{Martin-Olalla2018}, helped by artificial light. People find relief in a potential daily benefit: leaving work occurs more frequently before sunset. In that circumstances, DST ---yet another advance--- is less attractive. Quite often, this setting is usually achieved by other means: by extending DST to the winter (permanent DST). 

    \begin{figure*}
     \centering
     \includegraphics[width=0.9\textwidth]{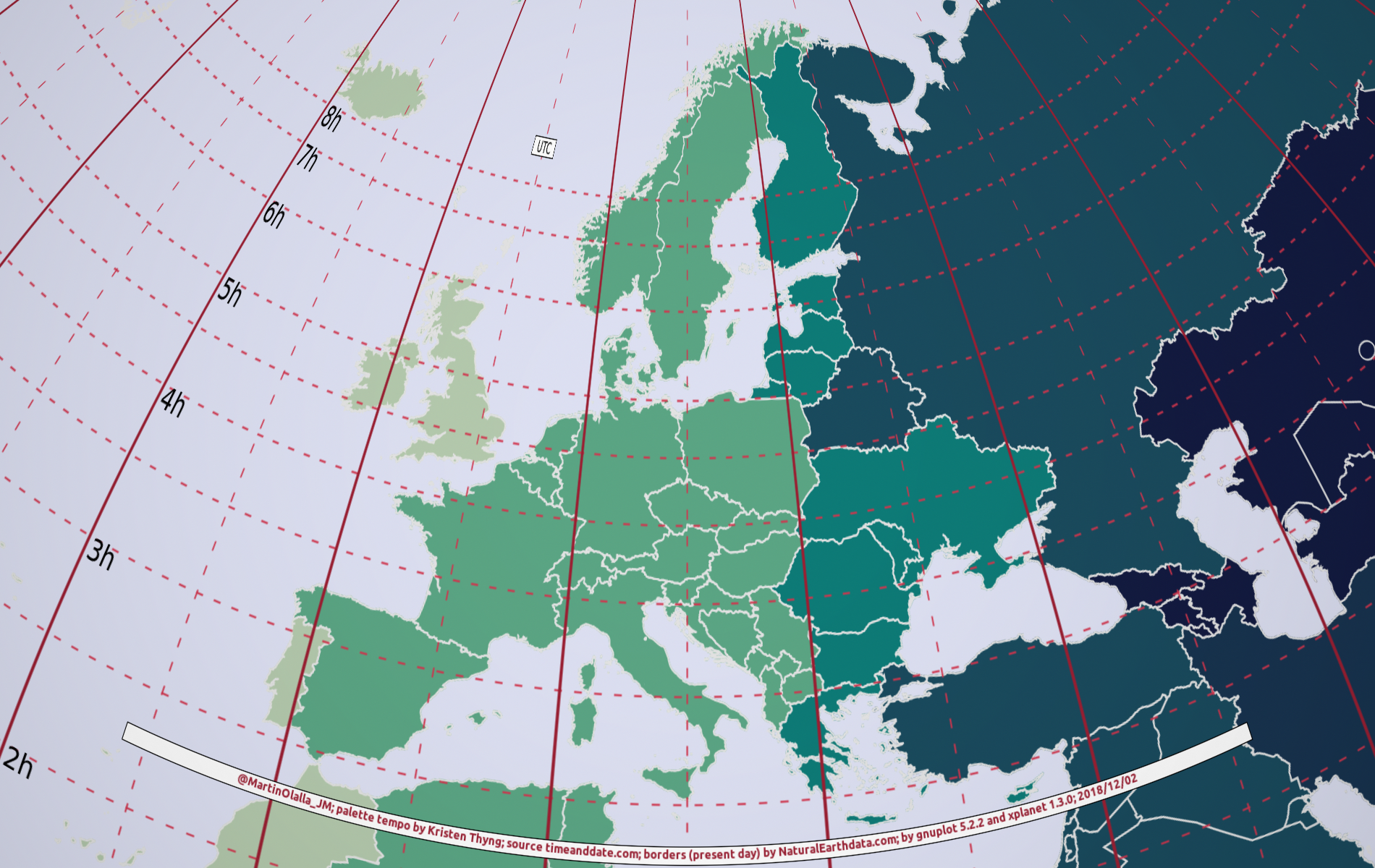}
     \caption{A map of Europe showing the time difference from winter sunrise/sunset to summer sunrise/sunset. Notice that if DST is onset then sunrise time difference decreases by one hour and sunset time difference increases by one hour as measured by local time.}
     \label{fig:e}
   \end{figure*}
    
   Looking in perspective to this process, the phase of the sleep/wake would remain nonseasonal. People would wake up quite after sunrise or quite before sunrise depending on season. They would be mere observers of the changes in the LD cycle.

   Finally at mid-latitudes DST efficiently binds sunrises, wake up times and clock time through seasons. It also offsets the winter lag. Notice that at this range people can not afford waking up long after sunrise in summer because later on, at noon, insolation will be high. Equally they find no relief in waking up long before winter sunrise, because winter photoperiod is longer, which easily helps increasing the shares of people leaving work before sunset.

   Although socio-cultural preferences are perhaps of the most importance to thoroughly understand this problem, the time difference between winter sunrise and summer sunrise ---see Figure~\ref{fig:e}--- is the quantity to address it technically. The impact of DST in all practical terms becomes more significant when the natural difference ranges from \SI{2}{\hour} to \SI{4}{\hour}, the impact is less significant above and below that range.

   Extra-tropical contemporary societies bound to clocks are doomed to a tricothomy. Either they face higher levels of solar exposition at noon in the hottest season of the year if the winter lag sustains year round. Or, they face an increasing activity before sunrise during the darkest season of the year if social timing is advanced relative to winter sunrise. Or they face DST, their transitions and their adaptations, which are temporary, and seasonally adapt the phase of human activity.

   A scientific understanding of the pros and cons of DST in terms of the circadian timing system, in terms of the public health, and in terms of the many socio-economic issues to which it is related must comprehensively address the consequences of every of these alternatives. My expert statement is quite simple: a social answer to this tricothomy depends significantly on latitude. Therefore it is unwise to end up with a binary statement: ``DST \emph{must/must not} be discontinued''.

Circles of latitude are characteristically different to each other. Europe spans a really wide range of circles of latitudes. That is why, I strongly and humbly suggest the European Commission to avoid looking for an all-in-one decision on this issue.

 \bibliographystyle{apsrev}

 \end{document}